\documentclass[twocolumn]{aastex61}
\usepackage{longtable}
\usepackage{float}

\usepackage{threeparttable}

\submitjournal{ApJ}

\shorttitle{Metallicities in M dwarf Planet Hosts}
\shortauthors{Wanderley et al.}

\begin{document}

\title{Metallicities of M Dwarf Planet Host Stars from Kepler, K2, and TESS observed by APOGEE: Trends with Exoplanetary Radii and Orbital Periods}

\correspondingauthor{Fábio Wanderley}
\email{fabiowanderley@on.br}

\author[0000-0003-0697-2209]{Fábio Wanderley}
\affiliation{Observatório Nacional/MCTIC, R. Gen. José Cristino, 77, 20921-400, Rio de Janeiro, Brazil}

\author[0000-0001-6476-0576]{Katia Cunha}
\affiliation{Steward Observatory, University of Arizona, 933 North Cherry Avenue, Tucson, AZ 85721-0065, USA}
\affiliation{Observatório Nacional/MCTIC, R. Gen. José Cristino, 77,  20921-400, Rio de Janeiro, Brazil}

\author[0000-0002-7883-5425]{Diogo Souto}
\affiliation{Departamento de F\'isica, Universidade Federal de Sergipe, Av. Marcelo Deda Chagas, S/N Cep 49.107-230, S\~ao Crist\'ov\~ao, SE, Brazil}

\author[0000-0002-0134-2024]{Verne V. Smith}
\affiliation{NSF’s NOIRLab, 950 N. Cherry Ave. Tucson, AZ 85719 USA}

\author[0000-0001-9205-2307]{Simone Daflon}
\affiliation{Observatório Nacional/MCTIC, R. Gen. José Cristino, 77, 20921-400, Rio de Janeiro, Brazil}

\begin{abstract}
One important property in studying the exoplanet population is the host star metallicity ([M/H]). In this study, we derived stellar metallicities and oxygen abundances for 48 M dwarf stars using the near-infrared high-resolution spectra from the SDSS APOGEE survey and synthetic spectra computed in LTE. We also derived and investigated the exoplanetary radii distribution for a larger sample of 246 exoplanets orbiting 188 M dwarf stars. The [M/H] versus [O/M] distribution obtained indicates that our sample is composed mainly of thin disk stars, which follow the behavior of the low-alpha sequence in the Milky Way thin disk. 
Small planets with radii smaller than 3R$_{\oplus}$ were found around stars with a range of metallicities (-0.6$<$[M/H]$<$+0.3), while larger planets of the sample orbit only stars with [M/H]$\geq0.0$. These results indicate that while small planets can form in different environments, larger planets preferentially form in metal-rich protoplanetary disks. Exoplanets with P$_{\rm orb}<$4.3 days orbit on average more metal-rich stars than planets with longer periods. This threshold is smaller than that found for FGK stars (8--10 days) and might be related to M dwarfs having a smaller dust sublimation radius. The distribution of exoplanets with R$_{\rm p}>$4R$_{\oplus}$ shows a concentration at orbital periods between 2 and 5 days, which may result from inward orbital migration. There is also a different behavior between single-detected exoplanets and planets from multiplanetary systems, with the latter being found on average around more metal-poor stars, and with planetary radii roughly up to 3 R$_{\oplus}$. 
\end{abstract}

\section{Introduction}

Deriving the fundamental physical properties of exoplanetary host stars (such as effective temperature, luminosity, surface gravity, stellar radius, or mass) is crucial for characterizing their exoplanetary properties in detail. One key property is stellar metallicity ([M/H]), which has been shown to influence exoplanet formation, architecture, and radius. 

Several works have shown that there are correlations between the metallicity of FGK host stars and the occurrence of different types of exoplanets. For example, giant planets tend to form around metal-rich stars \citep{fischer2005,santos2005,sousa2008,johnson2010,ghezzi2010,narang2018}, while smaller planets, such as super-Earths and sub-Neptunes are found around host stars having a wider range of metallicities \citep{buchhave2012,petigura2018}. Furthermore, there is also a dependence on the exoplanet's orbital period with metallicity. Hot small exoplanets (with radii less than $\sim$1.7 R$_{\oplus}$ and periods less than $\sim$8-10 days) orbiting FGK hosts seem to preferentially orbit more metal-rich stars compared to exoplanets having longer periods, for which the trend with metallicity is flat or negative (\citealt{mulders2016,petigura2018,wilson2018}; see also \citealt{teske2024}). \citet{ghezzi2021} in particular, derived metallicities for $\sim$800 Kepler F, G, K hosts from CKS and compared [Fe/H] distributions of stellar-hosts segregated into a variety of exoplanetary architectures, dividing the sample into 'hot' (P$_{\rm orb} < $10 days) and 'warm' (10$ < $P$_{\rm orb} < $100 days) systems for different exoplanetary sizes: super-Earths, sub-Neptunes, sub-Saturn, and Jovian exoplanets.  
They discuss several conclusions from these comparisons, but one general result found for the FGK stellar hosts is that the systems containing hot exoplanets tend to orbit more metal-rich stars when compared to stars hosting warm exoplanetary systems. In addition to orbital periods and exoplanetary radii, some studies have also investigated the relationship between stellar metallicities and exoplanet multiplicity for FGK hosts \citep{romero2018,weiss2018}. 

Although metallicities have been derived for large numbers of planet-hosting FGK stars, the number of accurate [M/H] determinations for the cooler and lower-mass M-dwarf planet hosts is much smaller. Even though the number of exoplanets detected around M dwarfs to date is considerably smaller than for the FGK stars, this number has been growing, particularly as a result of the TESS mission. Due to both their sheer numbers in the Galaxy and their importance as likely future planet host stars that will be targeted by such missions as the continuing JWST mission or future missions, such as, the Habitable Worlds Observatory (HWO) and PLATO. It becomes increasingly important to provide accurate determinations of the properties of M dwarf stars, especially their metallicities \citep{rojasayala2012,onehag2012,newton2014,maldonado2015,terrien2015,lindgren2016,lindgren2017,souto2020,souto2021,rajpurohit2018b,Passegger2022,birky2020,sarmento2021,marfil2021,lyer2023,bello2023,terese2025,behmard2025}.
However, there are significant challenges in determining stellar parameters and metallicities for cool stars and fewer works have investigated the metallicities for M dwarf hosts 
(see \citealt{schlaufman2010,mann2013b,neves2013,muirhead2014,gaidos2014,souto2017,souto2018,petigura2018,hirano2018,dressing2019,wang2024,hejazi2024,melo2024,gore2024,antoniadis2024}).

A large number of M dwarfs have been observed by the high-resolution near-IR SDSS APOGEE survey \citep{blanton2017,majewski2017_apogee} that has in its DR17 parameters and metallicities for tens of thousands of M dwarfs. However, since APOGEE surveyed primarily red giants in the Galaxy, the APOGEE Stellar Parameters and Abundances Pipeline (ASPCAP; \citealt{garciaperez2016_aspcap}) has been tailored to analyze red giants and is not optimized for the analysis of the spectra of M dwarfs, which are much more complex. Previous detailed analyses of APOGEE spectra used dwarf M members of open clusters as benchmarks to verify the accuracy of ASPCAP-derived metallicities, given that (apart from diffusion effects) members of an open cluster are expected to have homogeneous chemical abundances.
\citet{souto2021} analyzed a sample of F, G, K, and M dwarf star members of the open cluster Coma Berenices and showed that the coolest M dwarfs have ASPCAP metallicities that are lower by roughly 0.4 dex when compared to the metallicities of the warmer FGK star cluster members. Such metallicity offsets were later confirmed by \citet{wanderley2023} who studied Hyades M dwarfs covering a range in effective temperature from $\sim$ 3200 -- 3900 K, and finding a median metallicity for the cluster of 0.09$\pm$ 0.03 dex, which is in good agreement with literature results from the optical for solar-type and red-giant stars. ASPCAP metallicity results for the same M dwarf stars showed a trend as a function of the effective temperature of the star. Cooler stars with T$_{\rm eff} \sim$3200 K had systematically lower metallicities by up $\sim $0.3, which is around 0.4 dex smaller than the median metallicity that \citet{wanderley2023} obtained for the cluster. The systematic abundance offsets for M dwarf members of open clusters indicated that there are unresolved issues with the APOGEE ASPCAP metallicities, in particular for cool M dwarfs. Alternatively, the spectroscopic analysis methodology from the studies of Souto et al. has been well vetted to produce accurate metallicities for M dwarf stars.

\citet{wanderley2025} studied a sample of exoplanets having M-dwarf stellar hosts observed by APOGEE, focusing on the discussion of small (R$_{\rm p}<$4 R$_{\oplus}$) having orbital periods of P$_{\rm orb}<$100 days, to probe the exoplanet radius gap, R$_{\rm gap}$ (\citealt{fulton2017}) around cool, low-mass stars. This study detected a well-defined radius gap among the M-dwarf exoplanets at R$_{\rm gap}\sim$1.8$\pm$0.2R$_{\oplus}$, a value that was found to remain roughly constant over the orbital period range of P$_{\rm orb}\sim$1--20 days; this behavior of R$_{\rm gap}$ being $\sim$independent of P$_{\rm orb}$ is in contrast to that found for the hotter and more massive F, G, and K dwarf planet hosts, where the value of R$_{\rm gap}$ decreases with P$_{\rm orb}$ at a rate of d(logR$_{\rm gap}$)/d(logP$_{\rm orb}$)$\sim$ -0.09 to -0.11 \citep{vaneylen2018,martinez2019}.
The radius gap for exoplanets around M dwarfs, lies between the super-Earth and sub-Neptune exoplanetary regimes, and is likely sculpted by combinations of photoevaporation \citep{lopez2018} / planetary core-powered mass loss \citep{gupta2019}, with orbital migration (see also \citealt{gaidos2024,luque2022}). The differing behaviors of R$_{\rm gap}$ as a function of P$_{\rm orb}$ suggest differences in certain aspects of planetary formation and/or evolution around M dwarfs relative to more massive FGK dwarf stars.

The aim of this paper is to further probe the properties of M dwarf stars by exploring connections with exoplanet properties, such as exoplanet radius, orbital period, and system multiplicity. The first part of this paper investigates relationships of planet properties with host star metallicities and oxygen abundances, which were derived for a subsample of 48 M dwarfs, via quantitative spectroscopic analyses of APOGEE spectra. The second part of this paper analyzes a sample of 246 exoplanets that orbit 188 M dwarf stars (most of this sample was studied for the radius gap of small planets in \citet{wanderley2025}.
This paper is organized as follows. In Section 2, we present the sample of exoplanets and their host stars, along with the methodology for deriving stellar metallicities and oxygen abundances, and comparisons of these parameters with results from the literature. In Section 3, we discuss the results, which include the obtained stellar abundances, trends of exoplanetary radii and orbital period with host star metallicities, and trends of exoplanet radii with orbital periods focusing on large planets. Finally, Section 4 summarizes the conclusions.

\section{The Stellar and Exoplanetary Samples and Their Parameters}

The sample analyzed in this study comprises 246 exoplanets orbiting 188 M dwarf stars. This selection contains 218 exoplanets which were discussed in \citet{wanderley2025}, focusing on the study of the radius gap of small exoplanets (R$_{\rm p}<$4 R$_{\oplus}$). In this study, we extend the sample to include a population of larger exoplanets with the goal of investigating relations with orbital period for small and giant exoplanets. In addition, a subset of the M-dwarfs in our sample was observed by the APOGEE survey, allowing the determination of their stellar parameters (presented in \citealt{wanderley2025}), along with oxygen abundances, and metallicities (presented and discussed in this paper), as will be summarized below. The stellar (distances, effective temperatures, surface gravities, metallicities, projected rotational velocities, oxygen abundances, M$_{\rm K_{\rm s}}$ absolute magnitudes, and stellar radii) and planetary (transit depth measurements, planetary radii and orbital periods) parameters of our sample are presented in Table \ref{fulldata}.

\subsection{Stellar Parameters and Metallicities for the APOGEE Sample} 

\begin{figure}
\begin{center}
  \includegraphics[angle=0,width=1\linewidth,clip]{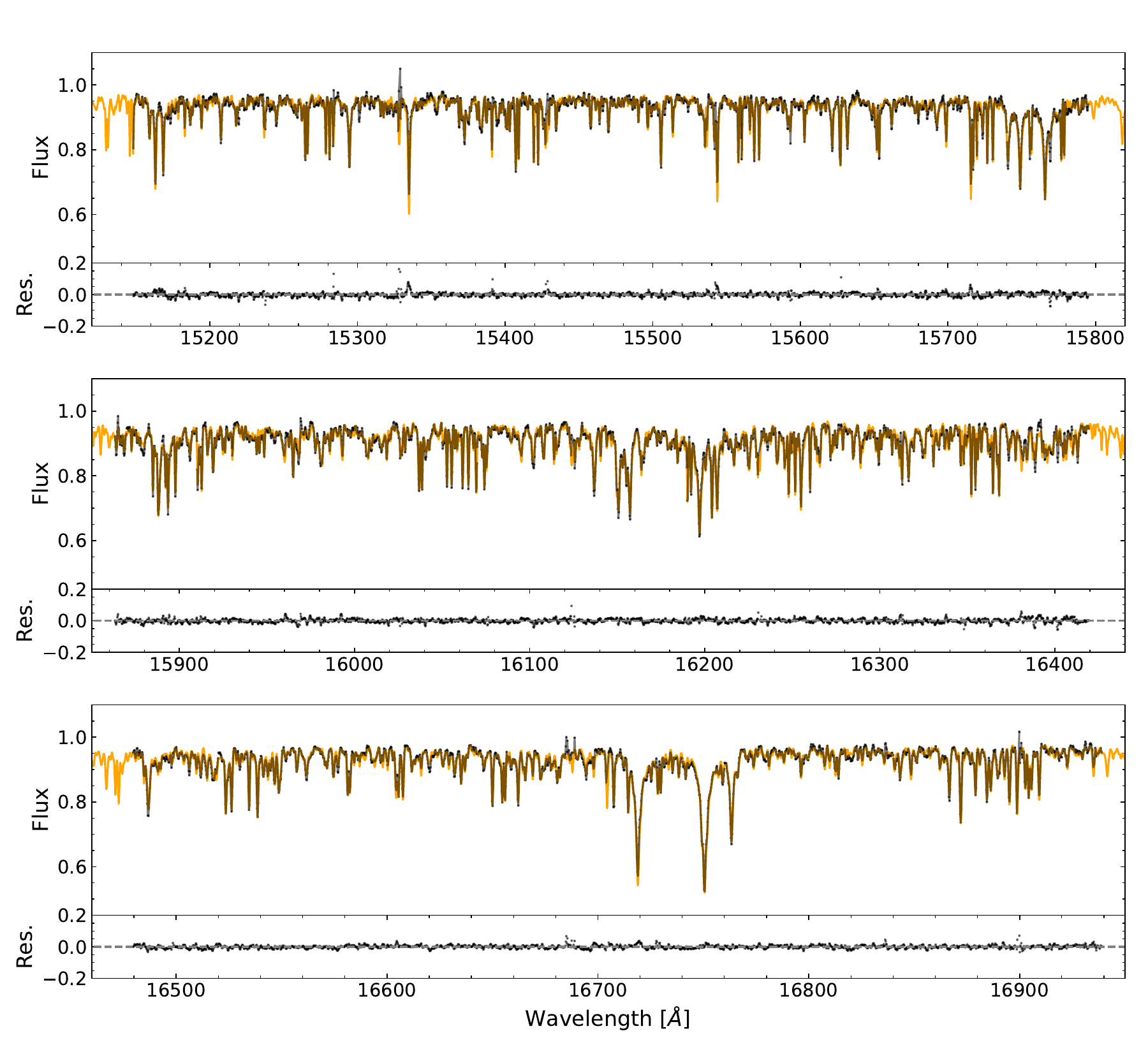}
\caption{Comparison between the APOGEE spectrum of the star Kepler-1308 (in black) and a spectral synthesis produced with the radiative transfer code Turbospectrum (in orange) for the derived stellar parameters for the star: T$_{\rm eff}$=3625 K, $\log{g}$=4.75, [M/H]=0.13, A(O)=8.75 and v$\sin{i}$=3.32 km s$^{-1}$. Each panel of the figure represents one of the three chips of the APOGEE detector. We also show the residuals between the observed spectrum and the best-fit synthesis at the bottom of each panel.}
\end{center}
\label{spec}
\end{figure}

The effective temperatures (T$_{\rm eff}$), surface gravities ($\log{g}$), and oxygen abundances (A(O)) for the APOGEE sample were derived by comparing spectral syntheses computed for different T$_{\rm eff}$ -- A(O) and $\log{g}$ -- A(O) pairs with the observed spectra, considering spectral windows containing water and OH molecular lines (the selected windows are listed in \citet{wanderley2025}. Metallicities ([M/H]) were derived through the overall best fits between synthetic and observed spectra. The projected rotational velocities (v$\sin{i}$) were determined using OH lines, similarly to the methodology discussed in \citet{wanderley2024}. 
The radiative transfer code Turbospectrum \citep{plez2012_turbospectrum} and MARCS model atmospheres \citep{gustafsson2008_marcs} were used for the computation of synthetic spectra.
As an example of the quality of the fits obtained, in Figure \ref{spec} we show the APOGEE spectrum of the M dwarf Kepler-1308 (in black), along with our best-fit spectral synthesis (in orange). 
The spectral synthesis was computed for T$_{\rm eff}$ = 3625 K, $\log{g}$ = 4.75, [M/H] = 0.13, A(O) = 8.75 and v$\sin{i}$ = 3.3 km s$^{-1}$. The three panels of Figure \ref{spec} correspond to the three APOGEE chips. We also show the residuals between the observed spectrum and our best-fit synthesis at the bottom of each panel.

\begin{figure}
\begin{center}
  \includegraphics[angle=0,width=1\linewidth,clip]{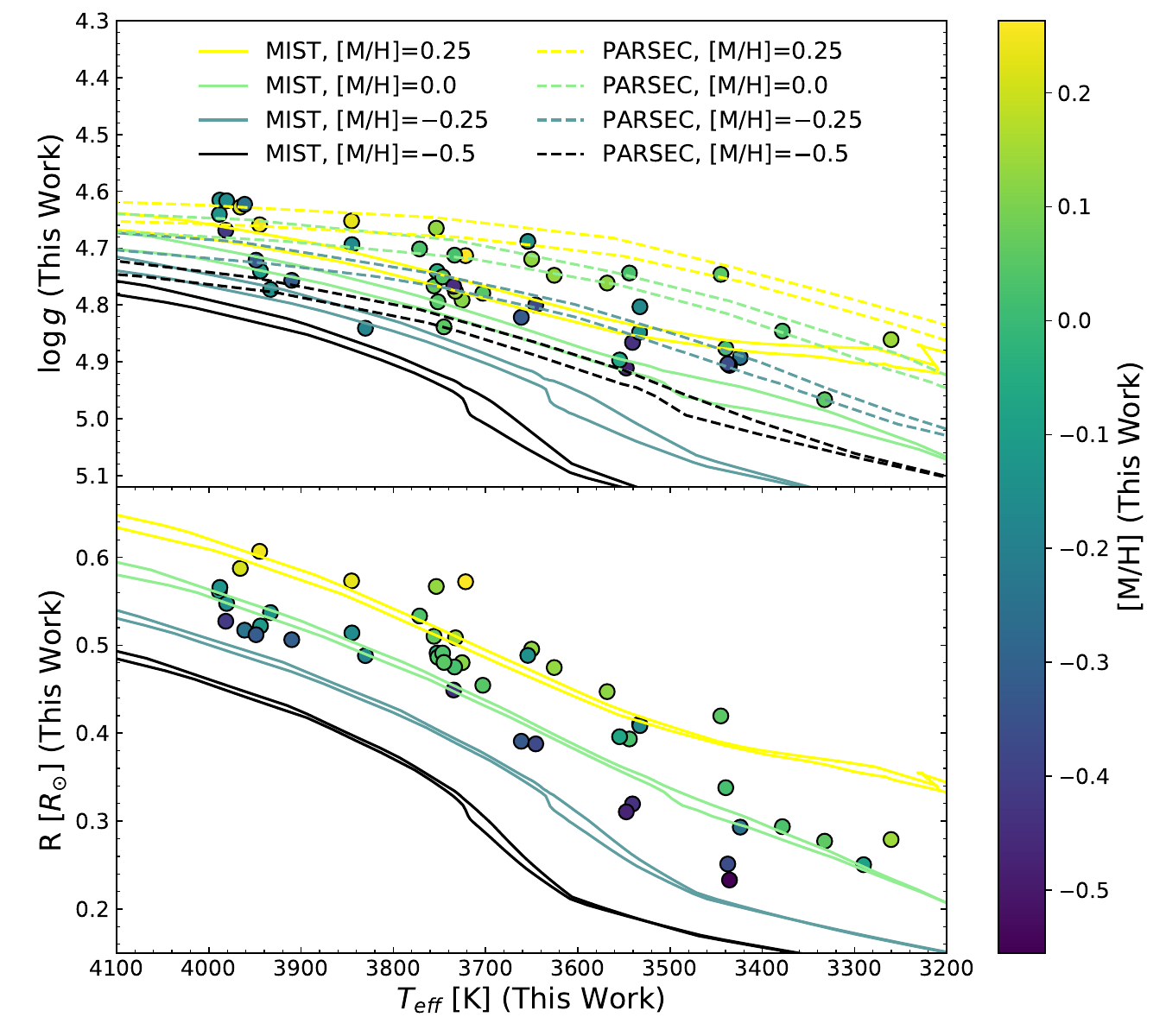}
\caption{The top and bottom panels show respectively the Kiel diagram and the relation between T$_{\rm eff}$ and stellar radii for the APOGEE sample M dwarfs. Both panels show MIST isochrones with ages of 1 Gyr and 10 Gyr and metallicities of $-$0.50, $-$0.25, +0.00, and +0.25. The top panel also shows, for comparison, PARSEC isochrones, with the same ages and metallicities chosen for the MIST isochrones.}
\end{center}
\label{kiel}
\end{figure}

Figure \ref{kiel} provides a visual summary of the stellar parameters for the APOGEE sample M dwarfs, both in a Kiel diagram, with $\log{g}$ versus $T_{\rm eff}$ (top panel), and in the stellar radius as a function of $T_{\rm eff}$ (bottom panel). In both panels, results are shown as a function of the derived stellar metallicity, according to the color bar on the right side of the figure. The individual ages of the studied M dwarfs are unknown, but for reference, we also show sets of MIST isochrones \citep{choi2016_MIST} for ages of 1 Gyr and 10 Gyr and metallicities of $-$0.50, $-$0.25, +0.00, and +0.25 dex. It is clear from the isochrone locations that the metallicity effect is more important than the age effect, as there are only small shifts between the 1 and 10 Gyr isochrones (they are adjacent to each other), while the shifts are much larger between the isochrones having different metallicities. For comparison, PARSEC isochrones \citep{Bressan2012_parsec} for the same metallicities and ages as the displayed MIST isochrones are added to the top panel of Figure \ref{kiel}, and it becomes evident that there are systematic differences between the predictions from the MIST and PARSEC isochrones.

In general, the $\log{g}$, stellar radius, and effective temperature results in this study follow a similar trend/slope of $\log{g}$ and radius versus $T_{\rm eff}$ as predicted by the models, with hotter M dwarf stars presenting on average larger radii and smaller $\log{g}$. 
Our $T_{\rm eff}$ versus $\log{g}$ results tend to fall above the MIST isochrones corresponding to their metallicities, while for the PARSEC isochrones they tend to fall roughly within range.
Most of the results for stellar radii versus $T_{\rm eff}$ (bottom panel of Figure \ref{kiel}) fall roughly within the MIST isochrone ranges for the solar and metal-rich isochrones, while, as will be shown in Section 3.1, our M dwarf sample has a significant number of metal-poor stars, with [M/H] between 0 and -0.6. This apparent mismatch between the derived stellar radii (and given metallicities) and those predicted by the isochrones is a well-known fact for M dwarfs \citep{reiners2012,Jackson2016,Jackson2019,jeffers2018,Kesseli2018} that can be explained as the result of radius inflation caused by magnetic fields \citep{wanderley2023}. (See also \citealt{wanderley2024b} that derived mean magnetic fields for many M dwarf hosts in common with this sample, and found mean magnetic fields ranging from $\sim$0.2 to $\sim$1.5 kG).


\subsubsection{Comparisons with the Literature}

\begin{figure}
\begin{center}
  \includegraphics[angle=0,width=1\linewidth,clip]{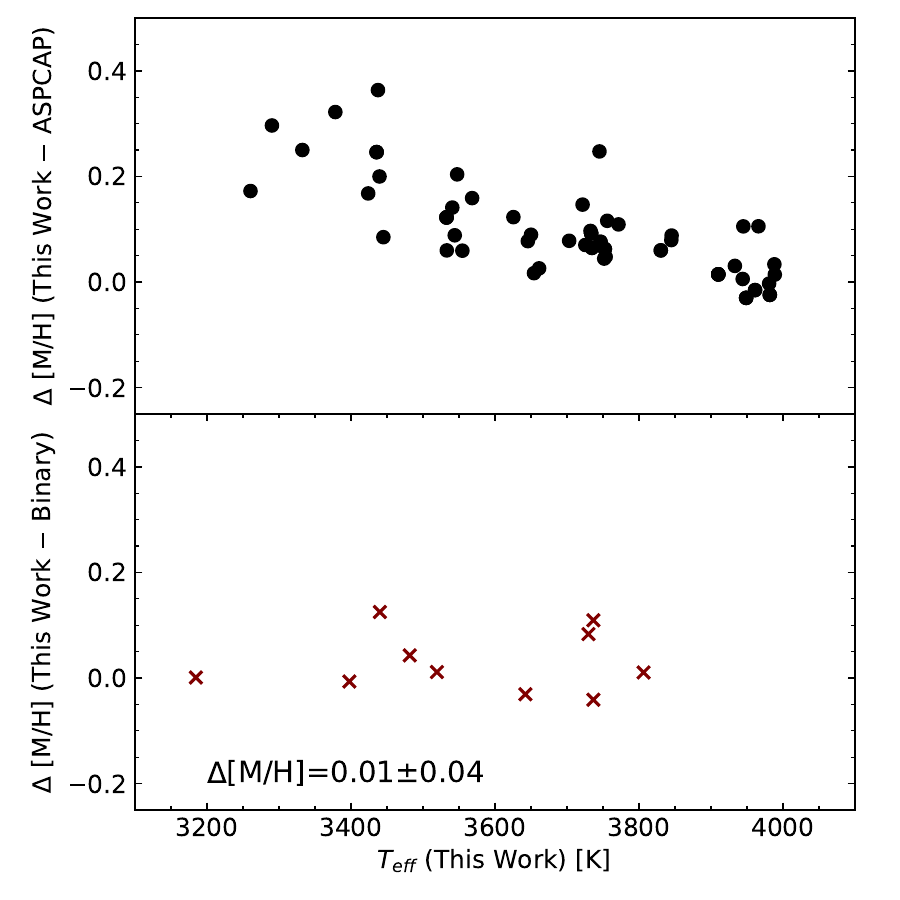}
\caption{Upper panel: a comparison between the derived metallicities for the APOGEE sample M dwarfs with results from ASPCAP DR17, as a function of effective temperatures derived from this work. Lower panel: a comparison between the derived metallicities for a sample of M dwarfs with FGK companions and the results for these companions from the literature, as a function of effective temperatures derived from this work.} 
\end{center}
\label{deltamet}
\end{figure}

One of the main goals of this study is to investigate possible trends of the metallicities of M dwarf planet hosts and their exoplanets and, to do so, we need a reliable metallicity scale for the M dwarfs.  A comparison between our metallicities with the ASPCAP pipeline \citep{garciaperez2016_aspcap} DR17 metallicities is shown in the upper panel of Figure \ref{deltamet}. There is a clear trend of the metallicity difference with T$_{\rm eff}$, with $\Delta$ [M/H] ``This Work $-$ ASPCAP" reaching $>$0.3 dex for the coolest stars in our sample, while the offset in metallicity is small for the hottest M dwarf stars with effective temperatures $\sim$3900 K. 
Systematic offsets in the ASPCAP metallicities of cool M dwarfs are well known by the APOGEE ASPCAP team, given that the ASPCAP pipeline was optimized for the analysis of red-giant spectra. As mentioned in the Introduction, metallicity discrepancies have been reported in the literature \citep{wanderley2023,souto2021} for benchmark M dwarf stars from open clusters. 

\textit{Benchmark Stars}:
To further test our metallicity scale, in this study, we also derived stellar parameters and metallicities for 10 benchmark M dwarfs previously studied in \citet{souto2020,souto2022}, which are binary stars that have FGK companions. These results are presented in the lower panel of Figure \ref{deltamet}. We show the difference, as a function of T$_{\rm eff}$, between our metallicities for the M dwarfs and the metallicities of the warmer companions that were collected from other works in the literature (\citealt{adibekyan2012,bensby2014,carretta2013,desilva2015,ghezzi2010,lambert2004,mann2013a,mishenina2008,ramirez2007,ramirez2012,reddy2006}). We find for this comparison a median offset of $+0.01 \pm 0.04$ between the M dwarf and the FGK star metallicities, which is a small difference and generally in line with the assumption of chemical homogeneity of stars born in the same molecular cloud. Different from what is observed in the upper panel of Figure \ref{deltamet}, the metallicity differences when using as comparisons results from high-resolution optical studies for warmer stars (bottom panel) do not show a clear trend with the effective temperature, suggesting that our metallicity scale are reliable. 

\subsection{Stellar and Planetary Radius Determinations} 

We used the M$_{\rm K_{\rm s}}$ versus stellar radius (R${*}$) relation in Equation \ref{rmk} derived in \citet{wanderley2025} to determine stellar radii for an additional 24 M dwarf stars that were not analyzed in their work.

\begin{equation}
    R_{*}=a_{0} + a_{1} M_{\rm K_{\rm s}} + a_{2} M_{\rm K_{\rm s}}^{2}
    \label{rmk}    
\end{equation}

\noindent where a$_{0}$=1.7420, a$_{1}$=$-$0.2925, and a$_{2}$=0.0123. 
To keep all results on the same scale, we used M$_{\rm K_{\rm s}}$ absolute magnitudes from 2MASS K$_{\rm s}$, and stellar distances from \citet{bailerjones2021}. The K$_{\rm s}$ magnitudes were de-extinct using dust maps from \citet{green2018}.
We used transit depth measurements compiled from the literature to determine exoplanetary radii for 28 planets using Equation \ref{eqtransit} below:

\begin{equation}
    R_{p}= \Delta F^{0.5} \times R_{*}
    \label{eqtransit}    
\end{equation}

\noindent where $\Delta$F is the transit depth, given by the ratio between the cross-sections of the planet and the star.

\section{Discussion}

\subsection{Derived Metallicities and Oxygen Abundances of M Dwarf Planet hosts in Galactic Context}

Stellar metallicities and oxygen abundances derived for the M-dwarf planet hosts are used to place this sample of stars within the context of Galactic chemical evolution.
Massive stars (M$\ge$ 8--10 M$_{\odot}$), which end their evolution as core-collapse supernovae (SNII), synthesize large amounts of $\alpha$-elements (such as O and Mg), while mass-accreting white dwarf members of binary systems that exceed the Chandrasekhar mass limit explode as supernovae of Type Ia (SNIa) and produce significant quantities of iron-peak elements (such as Fe or Ni).
Since white dwarfs result from the evolution of relatively low-mass stars, in contrast to the high-mass stars that undergo core collapse, it requires longer timescales for the nucleosynthetic products from SN Ia to contribute to the build-up of elements in the Galaxy when compared to those elements synthesized and ejected from SN II. 
Due to the different evolutionary timescales between SNII and SNIa, the elemental abundance relation between [$\alpha$/M] and [M/H] will evolve over time, reflecting the balance between SNII and SNIa in driving the chemical evolution of the Galaxy.

\begin{figure}
\begin{center}
  \includegraphics[angle=0,width=1\linewidth,clip]{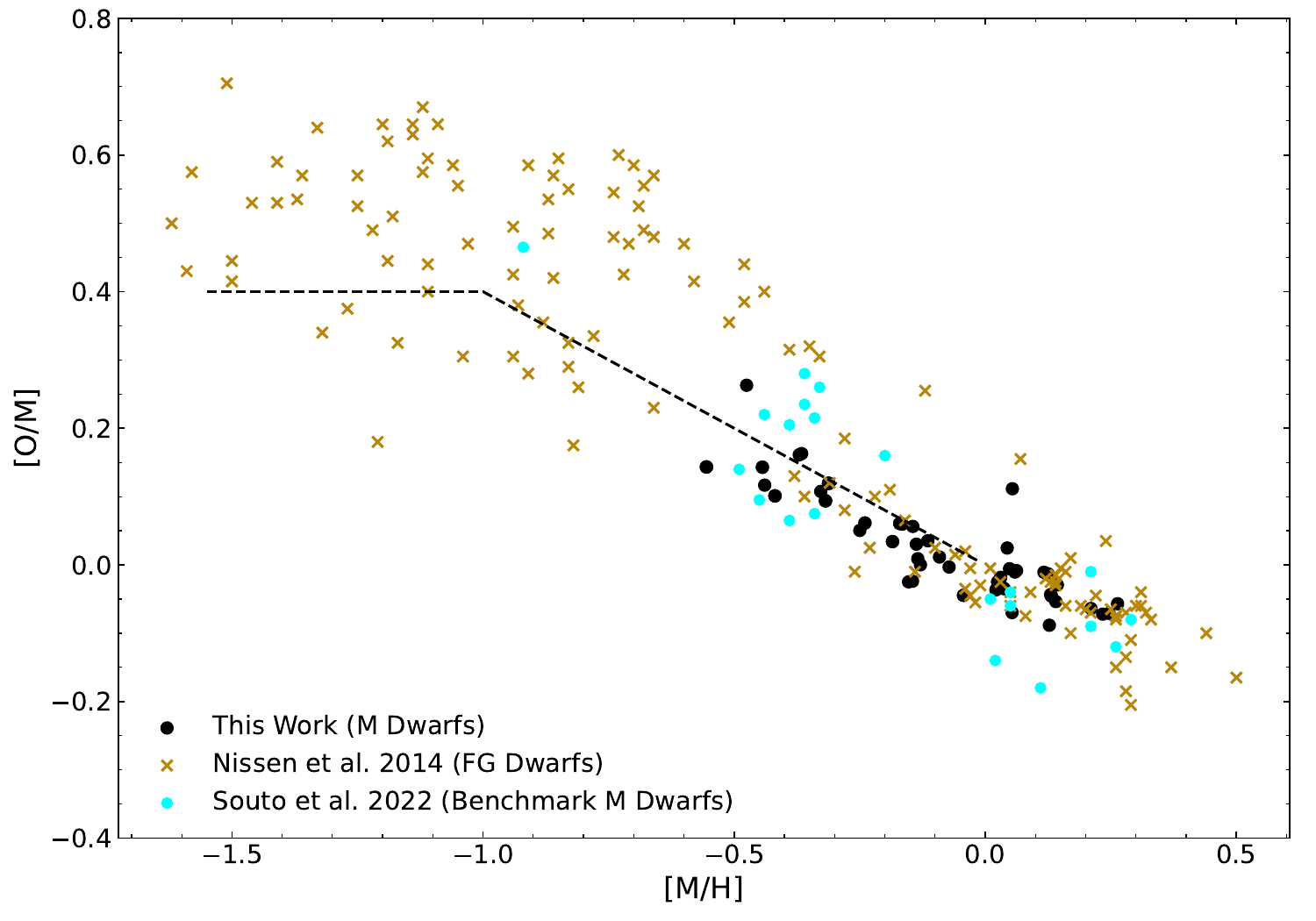}
\caption{Distribution of [O/M] as a function of metallicities. The studied sample of M dwarf planet hosts is shown as black circles, benchmark M dwarfs from \citet{souto2022} are represented as cyan circles. Results for F and G main-sequence stars from the high-resolution optical study of \citet{nissen2014} are also shown as orange x's. The agreement between our results and the optical results is very good, and defines well the thin disk trend. The black dashed line represents an expected schematic relation between both variables in the Milky Way galaxy.}
\end{center}
\label{galevo}
\end{figure}

The relation of [O/M] versus [M/H] for M-dwarf planet hosts from our sample is presented in Figure \ref{galevo} (as black circles). Our results define a tight relation of [O/M] versus [M/H] with small scatter that follows the Galactic thin-disk trend. As stars become more metal-poor, their [O/M] ratios increase as [M/H] decreases with a slope of $\sim$-0.4 until [M/H]$\sim$$-$1, where values of [O/M] reach a plateau corresponding to the ratio of O/M in SN II yields. This behavior is illustrated by the black dashed line in Figure \ref{galevo}, which represents a schematic relation between oxygen and metallicity of [O/M]= $-$0.4$\times$[M/H] down to [M/H]=$-$1.0, below which the plateau value of [O/M]=+0.4 is reached (we note that this particular chemical abundance pattern is shown as it is in the input abundances for the MARCS stellar model atmosphere grid \citealt{gustafsson2008_marcs}) used in this study. 

An independent comparison to our derived M dwarf abundances is shown for the detailed high-resolution optical study of F and G main-sequence stars (from \citealt{nissen2014}; orange x's). Our [O/M] - [M/H] values align extremely well with the low-$\alpha$ sequence thin disk stars from the \citet{nissen2014} sample. 
Two stars in our sample (K2-72 and LHS-1815) have elevated ratios of [O/M] when compared to the other M dwarf stars at their metallicities, and these are potentially part of the high-$\alpha$ sequence disk. It is interesting  that the high-$\alpha$ M dwarf star with near-solar metallicity (LHS 1815), falls near a star from the \citet{nissen2014} sample, HD 168443, which is classified as a thick disk star in that study based on its kinematics. 
We used the astroNN python package \citep{leung2019} to derive kinematics for these two stars and found that LHS-1815 has a total space velocity with respect to the LSR of 114 km/s, identifying it as a probable thick disk star, while K2-72 has a space velocity of 67 km/s, placing it near the boundary between the thin and thick disk populations.
Figure \ref{galevo} also shows the results for benchmarks from \citealt{souto2022} (cyan circles). Our [M/H] versus [O/M] abundances align well with their results.
Their sample consists of M dwarfs with angular diameters measured by interferometry, as well as binary M dwarfs with a solar-type companion. That study also analyzed APOGEE spectra and a spectral analysis methodology that is similar to ours, but their metallicities were based on Fe I lines and not determined from global fits like in this study (We recall that Figure \ref{deltamet} shows metallicity results for some of these benchmarks with our methodology). 

\subsection{Trends of Exoplanetary Radii and Orbital Period with Metallicity}

\begin{figure*}
\begin{center}
  \includegraphics[angle=0,width=0.7\linewidth,clip]{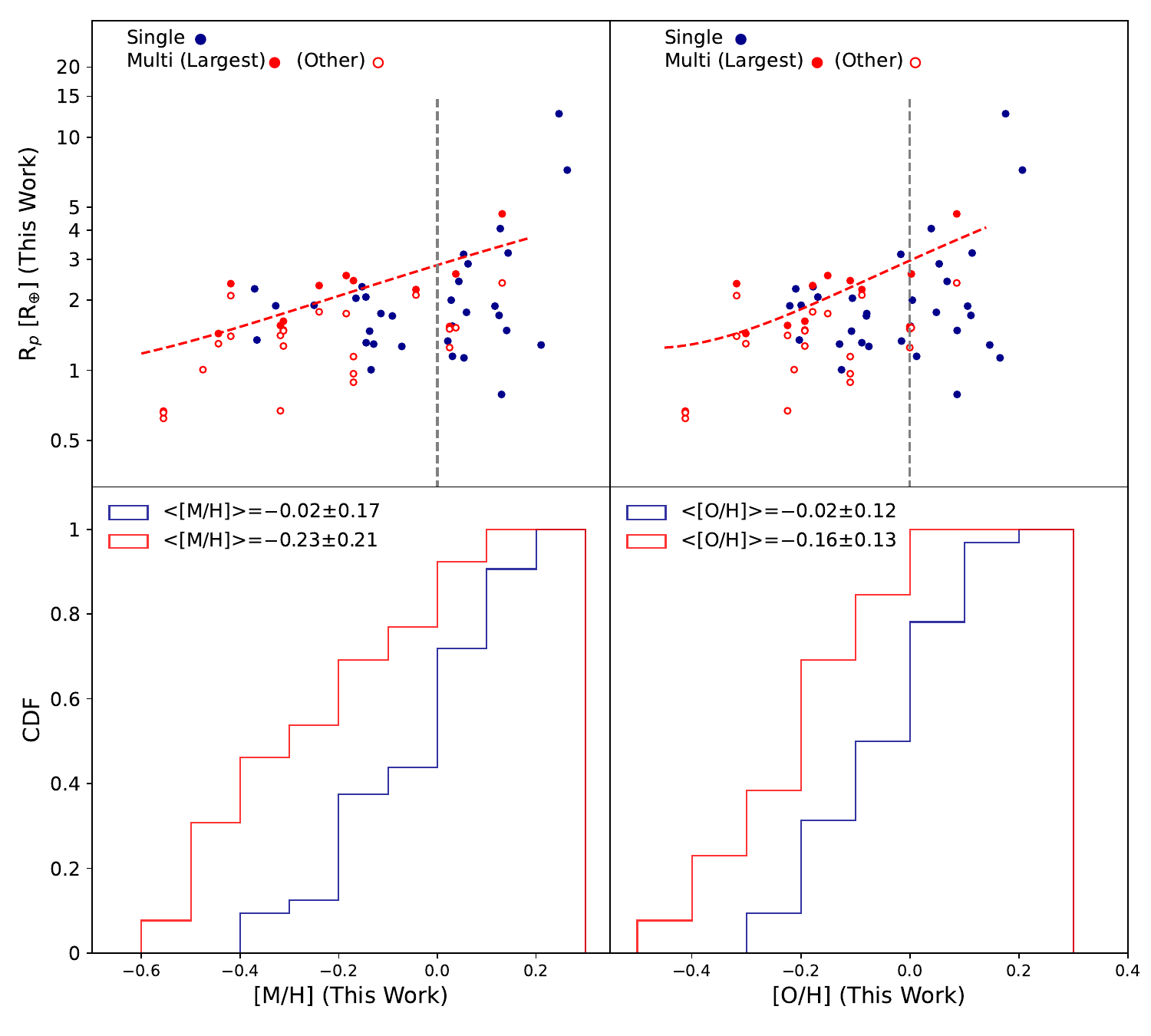}
\caption{Top Panels: Distribution of the derived planetary radii as a function of the derived host star metallicities (left panel) and [O/H] values (right panel). Single-detected planets are represented by filled blue circles, while systems known to have two or multiple detected exoplanets are represented by filled red circles (the largest exoplanet of each system) and red open circles (the remaining exoplanets of the systems). We also draw grey lines at solar metallicity ([M/H]=0) and [O/H]=0. The red dashed lines are quadratic fits between R$_{\rm p}$ and [M/H] (left upper panel) and [O/H] (right upper panel) considering the largest exoplanets of the multiple systems. Bottom Panels: Cumulative distributions of metallicities (left panel) and [O/H] (right panel) for singles (in blue) and multis (in red). We also show the mean$\pm$STD metallicities for each distribution.
}
\end{center}
\label{rplmet}
\end{figure*}

Figure \ref{rplmet} shows the exoplanetary radii as a function of metallicity (left panel), as well as [O/H] (right panel) for the studied sample of M-dwarf host stars. 
The 65 planets orbiting 45 M dwarfs displayed in Figure \ref{rplmet} have been identified as being in single (blue filled circles) or multis (red circles), with the largest known exoplanets in multi-planetary systems shown as filled red circles, while the remaining exoplanets from the multi-systems are shown as red open circles. 
The behavior of R$_{\rm p}$ with host star metallicity is similar for both the metallicity [M/H] and the oxygen abundance [O/H], both showing an overall positive trend of increasing R$_{\rm p}$ with [M/H] and [O/H], with the two highest metallicity M dwarfs hosting the two largest planets in our sample: Kepler-45b (hot Jupiter, R$_{\rm p}$ = 12.6 R$_{\oplus}$), host star metallicity of [M/H]=+0.25; and Kepler-1624 b (sub-Saturn, R$_{\rm p}$=7.2 R$_{\oplus}$), host star metallicity of [M/H]=+0.26.
The vertical dashed line marks the solar metallicity, and we can see that the metal-poor M dwarfs ([M/H]$<$0) host only smaller planets, with R$_{\rm p}$ less than approximately 2.5 -- 3 R$_{\oplus}$, while the metal-rich M dwarfs ([M/H]$>$0) host planets that span the entire range in planetary radii of our sample, with giant planets orbiting more metal-rich hosts. (This is also seen in the \citet{hirano2018} literature compilation of heterogeneous metallicities). 
The signature found here for M dwarfs is generally in line with what has been previously found for F, G, and K hosts which showed that Neptune size and smaller exoplanets orbit stars with a range of metallicities, while larger exoplanets tended to be found preferentially around stars with [M/H]$\geq$0 \citep{buchhave2012,sousa2018,petigura2018,ghezzi2010,adibekyan2019,ghezzi2021}. The general increase in R$_{\rm p}$ with [M/H] (or [O/H]) is found for both single planets, as well as the largest planets in multi-systems.  
For exoplanets that are the largest members of multiple systems, a quadratic relation of the form: R$_{\rm p}$= 2.83 + 4.20 [M/H] + 2.42 [M/H]$^{2}$ (in R$_{\oplus}$ units, shown by the red dashed line in Figure \ref{rplmet}), quantifies the positive correlation between planetary radius and metallicity of the host star. A similar quadratic fit using oxygen abundances is also shown in the right panel for R$_{\rm p}$ versus [O/H]. The largest single planets for a given metallicity (upper envelope of the blue circles distribution) seem to also follow the same relation as that defined by the largest planets in multis. Notably, there are no planets lying significantly above the red dashed curves and in particular, there is an absence of exoplanets with R$_{\rm p}$ $\gtrsim$ 2.5 R$_{\oplus}$ in our sample around stars with [M/H] $<$ 0. 
This may indicate a threshold in the maximum planet size with metallicity. Such behavior is also generally seen for the F, G, K hosts (see Figure 3 in \citealt{ghezzi2021}), although the threshold in that case was taken as 4.4 R$_{\oplus}$. 

A feature of Figure \ref{rplmet} is the different metallicity distributions between the single and multi-systems: of the 32 single system planets, 14 orbit M-dwarf hosts with [M/H]$<$0.0 and 18 have hosts with [M/H]$>$0.0, while for the 13 multi-systems, the respective numbers are 10 and 3. The statistics remain similar if oxygen is used as the metallicity indicator, with the breakdown for single systems being 16 ([O/H]$<$0.0) and 16 ([O/H]$>$0.0), and 11 and 2, respectively, for multis.
This difference in metallicity between single and multi-system M-dwarf hosts can be quantified by mean differences, with the mean$\pm$STD stellar metallicity for multi-planet and single-planet systems being, respectively, $-0.23 \pm 0.21$ and $-0.02 \pm0.17$, indicating an offset in metallicity between the two populations. 
A cumulative metallicity distribution between singles and multis is shown in the bottom panels of Figure \ref{rplmet}). It is clear that the metallicity distribution of singles versus multis is quite different. 
Using a two-sample Kolmogorov-Smirnov (K-S) test, we obtain a p-value of 0.002, which indicates at a 99$\%$ confidence level that the metallicity distributions of multi and single samples are likely not drawn from the same population.

Metallicity differences between single- and multi-planet systems around M dwarfs have also been pointed out by \citet{rodriguez2023}, using heterogeneous metallicity values from the NASA Exoplanet Archive. Such metallicity effect for M-dwarf hosts is in contrast to the lack of differences in metallicity between single and multi-systems for the hotter (T$_{\rm eff}\sim$ 4600 - 6500 K) and more massive F, G and K planet hosts from \citet{romero2018, weiss2018, ghezzi2021}.  
In comparisons between singles and multis, it should be pointed out that near-coplanarity is a requirement for the detection of multiple planets via transits, so the differences in [M/H] distributions between multi and single systems could be that a significant fraction of the single systems may be, in fact, multi-systems that are simply less coplanar than the observed metal-poor multi-systems.  A reason for such differences in planetary orbital inclinations could be that metal-rich proto-planetary disks enhance the occurrence rates of distant large planets, which can then perturb the orbital planes of more closely orbiting planets; such a scenario is suggested by \citet{pan2025} and is shown in their figure 4 (right panel).  
Why such a metallicity-dependent process would lead to contrasting behavior between single- and multi-planet systems around F, G, K dwarfs when compared to the M dwarfs is not known, but it suggests differences in the formation and/or evolution of planetary systems around different planet-host stellar masses and luminosities.

\begin{figure}
\begin{center}
  \includegraphics[angle=0,width=1\linewidth,clip]{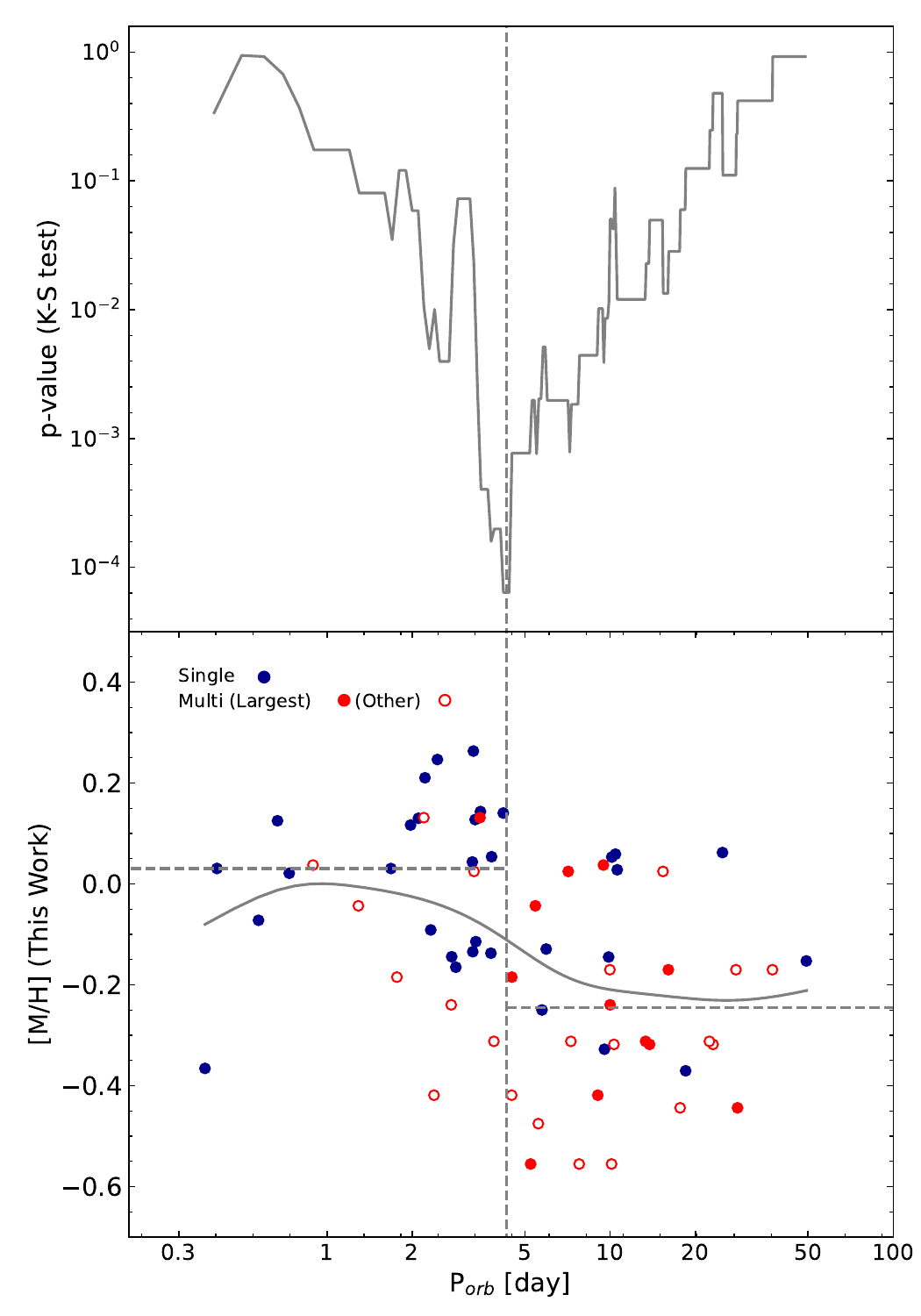}
\caption{Top Panel: p-values of the K-S tests calculated for the probability that the metallicity distributions of exoplanets having orbital periods above and below a given threshold in orbital periods are drawn from the same parent distribution. There is a statistically significant minimum of p-values (shown as the vertical dashed line) at P$_{\rm orb}$ =4.3 days for the M dwarf sample. Bottom panel: Distribution of the derived host M dwarf metallicities as a function of orbital periods. Blue circles are single-detected exoplanets, red filled circles are the largest exoplanets of each multiple system of the sample, and red open circles are the remaining exoplanets from these systems. The black curve shows the kernel regression of the mean metallicity. Horizontal lines indicate the median metallicity considering exoplanets with orbital periods below and above P$_{\rm orb}$ = 4.3 days (vertical dashed line).}
\end{center}
\label{metporb}
\end{figure}

We now investigate if there is a correlation between the metallicities of the M dwarf planet hosts and the exoplanet orbital periods, as was previously done for the F, G and K hosts. In the top panel of Figure \ref{metporb} we show the p-values from the Kolmogorov-Smirnov tests versus the exoplanet orbital period for our sample. To compute the p-values, we separated our sample considering different thresholds in orbital periods in a 0.1-day step. For each orbital period threshold, we divided the sample into a ``short-period" sample and ``long-period" sample, and we performed a K-S test between the two corresponding metallicity distributions, computing its p-value. 
The lowest p-value obtained was 6.4$\times$10$^{-5}$, corresponding to an orbital period of P$_{\rm orb}=$4.3 days (dashed vertical line in Figure \ref{metporb}). This p-value confirms with a 99$\%$ confidence level that the metallicity distributions for exoplanets with P$_{\rm orb}$ longer and shorter than 4.3 days are statistically different. 

The bottom panel of Figure \ref{metporb} shows the metallicities for the studied M dwarf hosts as a function of the exoplanet orbital period, keeping the same symbols as previously (blue circles for single exoplanets, filled red circles for the largest exoplanets of multiple systems, and open red circles for the remaining exoplanets of these systems). 
It is apparent from the distribution of the points in the figure that the metallicity of the host star and the exoplanet orbital periods may be anti-correlated, such that the host star metallicity is higher for exoplanets with shorter orbital periods, while it is lower for exoplanets with longer orbital periods. 
We did two non-parametric tests to assess if the exoplanet orbital period is anti-correlated with the host star metallicity. We computed the Kendall $\tau$ and the Spearman $\rho$ correlation coefficients for the sample, finding $-$0.26 (p-value=0.002) and $-$0.41 (p-value=0.001), respectively, indicating that the host star metallicities and exoplanetary orbital periods for this sample are anti-correlated. The low p-values found for these correlations are an indication that they are statistically significant. 

As discussed in previous studies of F, G and K hosts (e.g.,\citealt{wilson2018,mulders2016}) this could be related to protoplanetary disks of high-metallicity stars having higher amounts of solids, which could favor enhanced planetary formation closer to their host stars, or increase the efficiency of planetary migration in a dense protoplanetary disk \citep{adibekyan2013}.
This finding is also in agreement with \citet{beauge2013}, who suggests that exoplanets around metal-poor stars have delayed formation, presenting less inward orbital migration.

The bottom panel of Figure \ref{metporb} also shows the Kernel regression of the mean metallicity (gray curve) as a function of orbital period for our sample. This relation was derived using Equations 2 and 3 from \citet{wilson2018}, adopting a bandwidth of $\sigma$=0.50. The dashed horizontal lines in this figure are the median host star metallicities on each side of the threshold orbital period, with a median value above solar ($<$[M/H] $=$ +$0.03$) for P$_{\rm orb}$ $<$ 4.3 days and significantly below solar ([M/H] $=$ -0.24) for P$_{orb}$ $>$ 4.3 days. Both the Kernel regression and our median metallicities confirm that exoplanets with P$_{\rm orb}<$4.3 days preferentially orbit more metal-rich stars if compared to exoplanets with larger orbital periods.

Overall, such results for M dwarf hosts are similar to F, G, and K hosts in the sense that exoplanets with shorter periods tend to orbit higher-metallicity stars. \citet{wilson2018} found that the median metallicity of exoplanets with short periods in their FGK sample is +0.11 dex more metal-rich than the median metallicity for exoplanets having longer periods.
However, one significant difference between the results obtained here for the M dwarfs is that the orbital period threshold of 4.3 days implies a much smaller orbital separation when compared to the orbital separations of $\sim$8 -- 10 days obtained for FGK stars. These results were found for Kepler systems having F, G, and K hosts using metallicities from both APOGEE ASPCAP \citep{wilson2018}, LAMOST \citep{mulders2016,dong2018}, and CKS \citep{petigura2018}.

Differences in threshold orbital periods between the F, G, and K dwarfs and M dwarfs of, respectively, 8--10 days versus 4.3 days can be mapped into approximate differences in orbital sizes using typical masses for FGK dwarfs and M dwarfs.  Using a single mass of M$\sim$1.0M$_{\odot}$ for a G dwarf as an example,
and a threshold orbital period of P$_{\rm orb} \sim$ 9 days implies a semi-major axis, a$\sim$0.08 AU, while for the M dwarfs, a rough mass of M$\sim$0.4M$_{\odot}$ and a P$_{\rm orb}\sim$4.3 days leads to a$\sim$0.04 AU -- about a factor of two difference in the semi-major axes.  
One possibility to explain the threshold orbital period as a function of [M/H] that was explored, for example, by \citet{wilson2018}, is that this period occurs at a distance below which dust cannot form (a ``dust sublimation'' radius, R$_{\rm sub}$), which is a function of metallicity in the proto-planetary disk (e.g., \citealt{gabet2002}).  
Such a dust sublimation radius for M dwarfs is expected to fall at shorter orbital periods than for FGK stars, because R$_{\rm sub}\propto$(Stellar Luminosity)$^{1/2}$ \citep{gabet2002}. The dust sublimation radius hypothesis was discussed in \citet{wilson2018} but was discarded in their study of FGK hosts because they did not find a significant relation between the metallicity and the effective temperature of the host star in their sample. 
The lower orbital period for the M dwarfs obtained here in comparison with that for the FGK dwarfs, however, may be a signature of dust sublimation radius differences that appear due to the considerably larger contrast between the luminosities of the M dwarfs relative to the FGK dwarfs, than comparisons within the F, G, and K dwarfs themselves.

\subsection{Trends of Exoplanet Radii with Orbital Periods for Large Planets}

\begin{figure}
\begin{center}
  \includegraphics[angle=0,width=1\linewidth,clip]{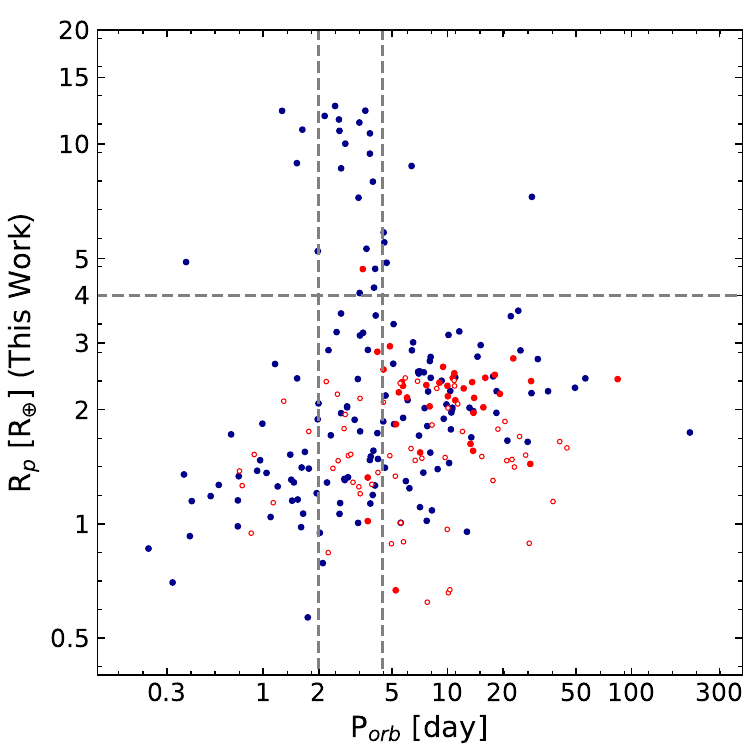}
\caption{Distribution of the derived planetary radii as a function of planetary orbital periods. Single-detected planets are the filled blue circles, exoplanets that are the largest planet of multiplanetary systems are represented by filled red circles, and the remaining exoplanet members of multiplanetary systems are represented by open red circles. A horizontal dashed line marks R$_{\rm p}$=4 R$_{\oplus}$, and the vertical dashed lines the 16$^{\rm th}$ and 84$^{\rm th}$ percentiles of the orbital period (at 2.0 and 4.4 days) distributions of exoplanets with R$_{\rm p}>$4 R$_{\oplus}$.}
\end{center}
\label{porbplot}
\end{figure}

The distributions of the exoplanet radius versus orbital period for the studied M dwarfs are shown in Figure \ref{porbplot}. (For a discussion of the small exoplanet sample in this figure and the radius gap, see \citealt{wanderley2025}). As in the previous section, we separate our sample into detected single exoplanets and exoplanets that are members of multiple systems. 

One of the main features of the planet radius vs orbital period diagram in Figure \ref{porbplot} is that M dwarfs, like their more massive FGK cousins, also host large exoplanets, including Jupiter-size planets, and that there is a marked tendency for such large planets to be both hot and to have orbital periods that fall within a small range. 
There is only one large planet in our sample having P$_{\rm orb}$ less than one day falling in the so-called Neptune desert previously noticed for FGK hosts, and there is only one large planet (R$_{\rm p}>$4 R$_{\oplus}$) with P$_{\rm orb}$ longer than 7 days. The concentration in orbital period is likely a consequence of inward orbital migration, leading to large planets born further from the snow line to be found in smaller orbital separations \citep{lin1996}. 
We computed the median of the orbital period distributions for exoplanets in our sample having R$_{\rm p} >$ 4 R$_{\oplus}$ and obtained 3.3$_{-1.3}^{+1.1}$ days (these limits are marked as vertical dashed lines in Figure \ref{porbplot} and represent the 16$^{\rm th}$ and 84$^{\rm th}$ percentiles of the distributions). 

The presence of hot Jupiters with P$_{\rm orb}$ $\sim$ 4 days is well known for exoplanets orbiting solar-type stars since early exoplanet detections based on RV measurements. In fact, the ``pile-up'' of large exoplanets around P$_{\rm orb}$=3--5 days is also observed for exoplanets around F, G, and K stars.
To make a comparison with what we find for the M dwarfs using more recent statistics, we compiled mean parameters for each exoplanet/stellar host from the NASA Exoplanet Archive separating the stellar hosts into spectral types F (6000 K $<$ T$_{\rm eff} <$ 7600 K), G (5200 K $<$ T$_{\rm eff} <$ 6000 K), and K (4000 K $<$ T$_{\rm eff} <$ 5200 K). 
We then derived the median $\pm$ percentiles of the orbital period distribution for exoplanets with R$_{\rm p} > $4 R$_{\oplus}$ of 4.5$_{-1.8}^{+34.8}$ days for F-type hosts, 5.0$_{-2.2}^{+46.6}$ days for G-type hosts, and 4.8$_{-2.1}^{+17.7}$ days for K-type hosts. These median orbital periods are generally in line and all fall roughly within the orbital period limits for our M-dwarf sample in Figure \ref{porbplot}. However, significant differences exist between the orbital-period distributions, where exoplanets orbiting M dwarfs are found to be much more confined in period, with less scatter, than the ones for exoplanets around F, G, and K stars, which extend to much longer orbital periods. Large exoplanets with P$_{\rm orb} >$ 5 days were barely detected around M dwarfs (2 planets out of 27 planets with R$_{\rm p}$ $>$ 4 R$_{\oplus}$). 

In contrast to the observed concentration of large planets in our sample, it is clear that the population of small exoplanets in Figure \ref{porbplot} spans a much wider range in orbital periods, showing no evidence of a ``pile up" at any orbital period. Another feature from Figure \ref{porbplot} is that all multis in our sample, except one, have R$_{\rm p} <$ 3 R$_{\oplus}$. This result is in agreement with what is found for solar-type in stars in the literature. \citet{dong2018} studied FGK hosts of exoplanets with orbital periods of up to 10 days using LAMOST data. They found that all hot Jupiters (R$_{\rm p}>$10 R$_{\oplus}$), and most Hot Neptunes (2 R$_{\oplus}<$R$_{\rm p}<$6 R$_{\rm oplus}$) from their sample are single-detected exoplanets. Figure 1 of their work shows that this signature is even cleaner if the threshold of R$_{\rm p}>$4 R$_{\oplus}$ is used, as we do here. This may suggest that multi-planetary systems might be less efficient in producing large exoplanets compared to single systems, but of course, it is also the case that there could be more exoplanets in singles that remain undetected. 
Unfortunately, given the unavailability of APOGEE spectra in this study, we are not able to derive metallicities for all exoplanet hosts in Figure \ref{porbplot}, but it is important to keep in mind that for our subsample of host stars/exoplanets with derived host metallicities (red circles in Figure \ref{metporb}), we find that the host stars of multis are generally more metal-poor compared to those of singles. This suggests that the metallicity and availability of solids in the protoplanetary disk may play a dominant role in the formation of large planets. We also note that the population of the largest exoplanets of multis has longer orbital periods than 3.5 days, with only four of these having shorter periods than 4.3 days. 

We note that in this study, we have combined results from three different missions (Kepler, K2, and TESS) and we investigated whether the overall conclusions would change if we had considered results from each mission independently. 
Segregating the sample results by mission would have the advantage of erasing possible systematic differences between the sets of results, but of course at the expense of smaller number statistics. Inspecting the three samples separately, we see that the TESS sample contains more exoplanets at shorter orbital periods ($P_{\rm orb}$$<$ 1 day) than the other two samples, and also contains a significant sample of giant planets.  Using the TESS sample alone, we recovered similar trends to those found in this study for the complete sample.

\section{Conclusions}
 
A sample comprised of 48 M dwarfs, was analyzed using high-resolution near-IR spectra from the SDSS APOGEE survey \citep{majewski2017_apogee} via spectrum synthesis computed in LTE. We used the Turbospectrum code \citep{plez2012_turbospectrum} and MARCS model atmospheres \citep{gustafsson2008_marcs} to derive stellar metallicities ([M/H]) and oxygen abundances ([O/H]). 
In addition, in this study, we examined the distribution of exoplanetary radii for a larger sample composed of 246 exoplanets orbiting 188 M-dwarf stars. This sample includes the exoplanets studied in \citet{wanderley2025}, which analyzed the radius gap of small planets around M dwarfs, and in this study, an additional 28 exoplanets were added to the sample.

The main conclusions from an analysis of the metallicities of host stars and their exoplanetary systems are:
\begin{itemize}
     \item Larger exoplanets, with R$_{\rm p}>3$ R$_{\oplus}$, are found only around the more metal-rich M dwarfs, with [M/H]$>$+0.0 (Figure \ref{rplmet}), suggesting that the formation of larger and more massive exoplanets requires more substantial reservoirs of ices and dust in the M-dwarf proto-planetary disks. Smaller exoplanets, with R$_{\rm p}<3$ R$_{\oplus}$, are found to orbit host stars that cover a wide range of metallicities (-0.6$<$[M/H]$<$+0.3; Figure \ref{rplmet}).
    
    \item A straightforward segregation of the transiting exoplanetary systems into those known to contain multiple exoplanets (multis) and those with only a single detected exoplanet (singles) reveals that the M dwarfs hosting multi-systems display a metallicity function that is shifted to significantly lower metallicities when compared to the single system [M/H] distribution (Figure \ref{rplmet} bottom panels). In contrast, the cumulative metallicity distribution functions for the multis versus singles for FGK hosts are indistinguishable \citep{weiss2018,romero2018,ghezzi2021}.
    In the literature, planetary-system models predict that systems that form from more metal-rich disks evolve into systems having planets with larger differences in mutual orbital inclinations than systems that formed from more metal-poor disks, which are more coplanar \citep{pan2025}. Increased mutual orbital inclinations with increasing [M/H] means that larger fractions of exoplanets in multi-systems will be undetected by transit observations, resulting in a larger fraction of apparently single systems orbiting higher-metallicity host stars. However, this does not explain the different behavior between the M dwarf and FGK hosts.
    \item There is an anti-correlation between [M/H] and orbital periods, such that shorter-period exoplanets tend to orbit more metal-rich M dwarfs, with a transition period between significantly different [M/H] distributions occurring at P$_{\rm orb}\sim$4.3 days.  Such an effect has been noted in the more massive and hotter FGK dwarfs, but with a significantly longer transition period of P$_{\rm orb}$=8--10 days \citep{mulders2016,wilson2018,petigura2018}. This transitional orbital period may be due to a ``dust sublimation radius", R$_{\rm sub}$, which has been discussed in the literature, with R$_{\rm sub}$ being the radius inside of which dust formation is inhibited by stellar flux.  The value of R$_{\rm sub}$ is proportional to L$_{*}^{1/2}$, so the differences between R$_{\rm sub}$ for M dwarfs compared to the FGK dwarfs are related to dust sublimation. 
    \item The chemical abundance pattern presented in Figure \ref{galevo} demonstrates that most of our sample of planet-hosting M-dwarf stars defines a tight sequence in [O/M] versus [M/H], which follows the general expectation of Galactic chemical evolution for the thin disk stellar population; this defines the low-alpha sequence. Two stars from our sample appear to possibly belong to the chemical thick disk. Future work with larger samples of M dwarfs will be able to explore possible differences between the [M/H] distribution functions for M-dwarf exoplanet hosts within the context of Galactic stellar populations.
\end{itemize}

The investigation of the exoplanet radii distribution as a function of orbital periods  for the entire sample resulted in the following conclusions:
\begin{itemize}
    \item Large exoplanets with R$_{\rm p}>4$R$_{\oplus}$ orbiting M dwarfs (Figure \ref{porbplot}) occupy a very narrow range in orbital period when compared to smaller exoplanets, with 21 of 27 large exoplanets falling within the period interval of P$_{\rm orb}\sim$2-5 days. This sharply-peaked ``pile-up'' of large exoplanets orbiting M dwarfs is in contrast to the orbital distributions for large exoplanets around the FGK dwarfs, which, although also orbited by a population of short-period large exoplanets, contain a significant fraction of giant planets orbiting well beyond 5-day periods, with P$_{\rm orb}\sim$5-100 days. Such short periods for large exoplanets are likely a result of inward orbital migration \citep{lin1996}, and the extremely narrow peak of large planet orbital periods around M dwarfs may indicate that orbital inward migration is more efficient for large exoplanets around these low-mass stars in comparison to solar-type stars.  
    \item All M-dwarf multi-systems in this sample, except for one, host exoplanets having R$_{\rm p}<$3 R$_{\oplus}$ and, since the transit-detected multis are more metal-poor than the single systems, this presumably reflects the lack of solids in metal-poor proto-planetary disks.
\end{itemize}

\startlongtable
\begin{deluxetable*}{lcccccccccccc}
\small\addtolength{\tabcolsep}{-2pt}
\tablenum{1}
\label{fulldata}
\tabletypesize{\scriptsize}
\tablecaption{Stellar and Planetary Data}
\tablewidth{0pt}
\tablehead{
\colhead{Hostname} &
\colhead{d} &
\colhead{T$_{\rm eff}$} &
\colhead{$\log{g}$} &
\colhead{[M/H]} &
\colhead{v$\sin{i}$} &
\colhead{A(O)} &
\colhead{M$_{\rm K_{\rm s}}$} &
\colhead{R$_{*}$} &
\colhead{Planet} &
\colhead{$\Delta$F} &
\colhead{R$_{\rm p}$} &
\colhead{P$_{\rm orb}$} \\
\colhead{...} &
\colhead{pc} &
\colhead{K} &
\colhead{...} &
\colhead{...} &
\colhead{km s$^{-1}$} &
\colhead{...} &
\colhead{...} &
\colhead{R$_{\odot}$} &
\colhead{...} &
\colhead{$\%$}  &
\colhead{R$_{\oplus}$}  &
\colhead{day}
}
\startdata
G 9-40 & 27.82$\pm$ 0.02 & 3423 & 4.89 & -0.25 & 5.36 & 8.46 & 6.97 & 0.30$\pm$0.01 & G 9-40 b & 0.34$\pm$0.03 & 1.90$\pm$0.12 & 5.75 \\
GJ 3470 & 29.38$\pm$ 0.02 & 3650 & 4.72 & 0.13 & 1.00 & 8.70 & 5.65 & 0.48$\pm$0.01 & GJ 3470 b & 0.60$\pm$0.02 & 4.06$\pm$0.14 & 3.34 \\
K2-25 & 44.62$\pm$ 0.06 & 3260 & 4.86 & 0.14 & 9.27 & 8.77 & 7.20 & 0.27$\pm$0.02 & K2-25 b & 1.16$\pm$0.06 & 3.19$\pm$0.21 & 3.48 \\
TOI-1685 & 37.6$\pm$ 0.02 & 3568 & 4.76 & 0.13 & 4.45 & 8.77 & 5.88 & 0.45$\pm$0.01 & TOI-1685 b & 0.13$\pm$0.01 & 1.72$\pm$0.08 & 0.67 \\
... & ... & ... & ... & ... &  ... & ... & ... & ... & ... & ... & ... & ... \\
\enddata
\begin{tablenotes}
        \small
        \item[] {\centering
        The T$_{\rm eff}$, $\log{g}$, and metallicity uncertainties are respectively $\pm$100 K, $\pm$0.20 dex, and $\pm$0.14 dex. The complete table is available in electronic format.\par}
\end{tablenotes}
\end{deluxetable*}

\acknowledgments

We thank the referee for comments that improved the paper. F.W. acknowledges support from PCI fellowship by MCTI. 
D.S. thanks the National Council for Scientific and Technological Development – CNPq. K.C. and V.S. acknowledge support by the National Science Foundation through NSF grant No. AST-2009507.

Funding for the Sloan Digital Sky Survey IV has been provided by the Alfred P. Sloan Foundation, the U.S. Department of Energy Office of Science, and the Participating Institutions. SDSS-IV acknowledges support and resources from the Center for High-Performance Computing at the University of Utah. The SDSS web site is www.sdss.org.
SDSS-IV is managed by the Astrophysical Research consortium for the Participating Institutions of the SDSS Collaboration including the Brazilian Participation Group, the Carnegie Institution for Science, Carnegie Mellon University, the Chilean Participation Group, the French Participation Group, Harvard-Smithsonian Center for Astrophysics, Instituto de Astrof\'isica de Canarias, The Johns Hopkins University, 
Kavli Institute for the Physics and Mathematics of the Universe (IPMU) /  University of Tokyo, Lawrence Berkeley National Laboratory, Leibniz Institut f\"ur Astrophysik Potsdam (AIP),  Max-Planck-Institut f\"ur Astronomie (MPIA Heidelberg), Max-Planck-Institut f\"ur Astrophysik (MPA Garching), Max-Planck-Institut f\"ur Extraterrestrische Physik (MPE), National Astronomical Observatory of China, New Mexico State University, New York University, University of Notre Dame, Observat\'orio Nacional / MCTI, The Ohio State University, Pennsylvania State University, Shanghai Astronomical Observatory, United Kingdom Participation Group,
Universidad Nacional Aut\'onoma de M\'exico, University of Arizona, University of Colorado Boulder, University of Oxford, University of Portsmouth, University of Utah, University of Virginia, University of Washington, University of Wisconsin, Vanderbilt University, and Yale University.

\facility {Sloan}

\software{Matplotlib (\citealt{Hunter2007_matplotlib}), Numpy (\citealt{harris2020_numpy}), Turbospectrum (\citealt{plez2012_turbospectrum})}

\bibliographystyle{yahapj}
\bibliography{references.bib}

\end{document}